# Single chip sensing of multiple gas flows

P. Bruschi, M. Dei
Dipartimento di Ingegneria dell'Informazione,
via G. Caruso 16,
56122 Pisa, Italy

M. Piotto
IEIIT Pisa - CNR,
via G. Caruso 16
56122 Pisa, Italy

*Abstract*-The fabrication and experimental characterization of a thermal flow meter, capable of detecting and measuring two independent gas flows with a single chip, is described. The device is based on a 4 × 4 mm$^2$ silicon chip, where a series of differential micro-anemometers have been integrated together with standard electronic components by means of post-processing techniques. The innovative aspect of the sensor is the use of a plastic adapter, thermally bonded to the chip, to convey the gas flow only to the areas where the sensors are located. The use of this inexpensive packaging procedure to include different sensing structures in distinct flow channels is demonstrated.

## I. INTRODUCTION

Detection and measurement of fluid flow is a crucial operation in many complex systems. In the field of automotive engines and household heating boilers, precise control of the combustion mixture allows efficiency improvement and reduction of pollutant emission. High resolution sensing of very small gas flows is an important target in the development of more effective ionic and cold gas thrusters for orbit and attitude control in scientific satellites [1]. Fine monitoring of multiple gas and liquid flow is also a requirement for controlling the correct evolution of chemical reaction. In particular, sensitive and miniaturized flow sensor are beginning to be integrated in microreactors [2,3], a new breed of silicon based devices devoted to handle small amount of reagents with strict temperature and pressure conditions.

It is interesting to observe that flow control is still preferably performed in an open loop fashion, i.e. relying on calibrated actuators such as, syringe [4] and peristaltic pumps [5], with no need of flow meters. The reason stands in the high cost, dimensions and insertion loss of traditional flow sensors.

In principle, such limitations could be overcome by using integrated thermal flow sensors [6], exploiting the advantages of silicon processing technologies to reduce costs, size and power consumption. Unfortunately, in spite of the huge variety of prototypes presented in the literature in last two decades, the number of commercial products currently available is still considerably small. One of the reasons is the necessity of developing proper packages, capable of protecting the fragile micromachined structures and, at the same time, allowing the interaction between the micrometric sensing structures and the gas. The method to convey the gas flow to the sensing structure without damaging the bonding wires and introduce as low as possible insertion losses has to be tailored to the particular application, taking into account specification of sealing, sensitivity, range and miniaturization [7]. Solutions based on placing the whole chip with its bonding wires inside the gas channel [8, 9] are not compatible with liquid flows or aggressive gases; furthermore, the channel cross-section cannot be reduced below several mm$^2$, so that, handling of very small flows is limited. On the opposite side, the alternative is placing the chip outside the flow channel, relying on a thermally conducting membrane or simply the pipe walls to allow heat exchange between the sensor elements and the fluid. [10, 11]. In this case, the need of heating a pipe portion several times longer than the chip size results in sensitivity reduction, higher power consumption and longer time transients.

Recently, we have proposed a packaging method based on a PMMA adapter, thermally bonded to the silicon chip [12]. The advantage of this approach is the possibility of conveying the gas flow to selected areas of a completely standard chip, with no need of particular bonding pad configurations, chip size or surface flatness.

In this work we describe the application of this inexpensive technique to the fabrication of a single chip flow meter capable of measuring two independent gas flows. The method is applied to a chip designed earlier than the development of the packaging method, so that the placement of the sensing structures was not optimized to include them in separate flow channels. Nevertheless, it will be demonstrated that the method can be applied to sense two gas flow using sensing structures separated one from the other and from the bonding pads only a few hundreds microns.

## II. DEVICE DESCRIPTION

The device is composed of two main elements: (i) a sensor chip and (ii) a plastic adapter used to convey the gas flow to the chip surface. The chip has been fabricated using the STMicroelectronics Bipolar-CMOS-DMOS process "BCD6" with 0.35 μm CMOS devices and three metal layers option. The flow sensing structures are classical thermal anemometers [1], formed by a heater (*p*–doped polysilicon resistor) and two temperature probes (thermopiles) placed upstream and downstream to the heater. Thermal insulation between the thermopiles and the heater is obtained by suspending the three elements over separate dielectric membranes by means of a standard micromachining procedure, briefly described in the next section.

A SEM (scanning electron microscope) micrograph of a sensing structure after the post-processing procedure is shown in Fig. 1, where the heater and thermopile conducting layers are not visible due to the upper dielectric layers.

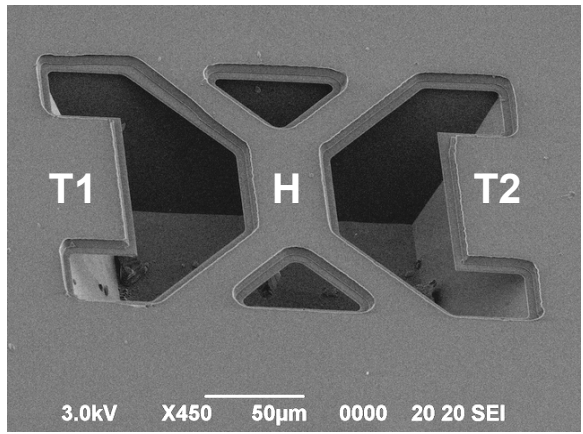

Fig.1. Scanning electron micrograph of a sensing structure after the post-processing procedure. T1,T2 and H indicate the thermopile and heater membranes, respectively.

The output voltage, proportional to the temperature difference between the downstream and upstream thermopile, is related to the flow velocity by a monotonic relationship over a wide velocity range. The sensing structures were equipped with three different thermopile types: *p*-polysilicon/Al (PA), *p*-polysilicon/*n*-polysilicon (PN) and n-polysilicon/Al (NA). Unfortunately most structures, designed using NA thermopiles, exhibited a near zero sensitivity, probably due to the low value assumed by the NA couple in the BCD6 process. On the other hand, the only two different structures, one of PA and the other of PN type, showed a high sensitivity; however, their placement was not favorable for the following packaging phase, due to their small spacing and vicinity of the bonding pad frame. The situation is clearly shown in Fig. 2, where the chip layout with the sensing structure is depicted.

The gas flow is conveyed to the micrometric sensing structures by means of the plastic adapter, applied to the chip surface in order to obtain sealed channels including the sensing structures.

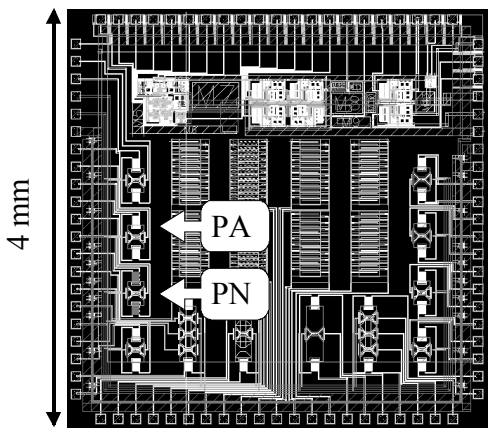

Fig.2. Placement of the PA and PP sensing structures on the chip layout.

The adapter face, which is placed into contact with the chip, is slightly smaller than the pad frame, so that it can be easily applied after bonding of the chip. Trenches of proper shape and position are milled on the adapter face, in order to make the gas flow interact with selected chip areas. The trenches are accessed by means of cylindrical holes perpendicular to the chip surface. The holes, that terminates on the opposite adapter face, can be connected to gas lines by means of capillary pipes. Sealing of the trenches to the chip surface is obtained using a thermal procedure explained in the next section. Two different configurations of the trenches were used in this work: the first one, shown in Fig. 3 (a) consists of two independent U-shaped trenches, including the PA and PP structure, respectively. The second configuration, shown in Fig. 3(b), consists of a single L-shaped trench, including both sensing structures. The first configuration represent the target two channel device while the second one, similar to the device presented in [12], is used for reference purposes.

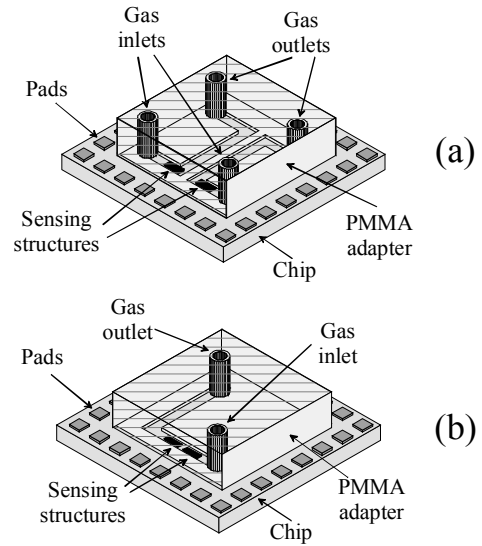

Fig.3 Perspective view of (a) the proposed device and (b) the prior version.

In both cases the trenches have a 0.5 mm × 0.5 mm rectangular cross section. The actual position of the trench on the chip surface in the case of Fig. 3 (a) is shown in Fig 4 (right) where the actual chip layout and dimensions are reported.

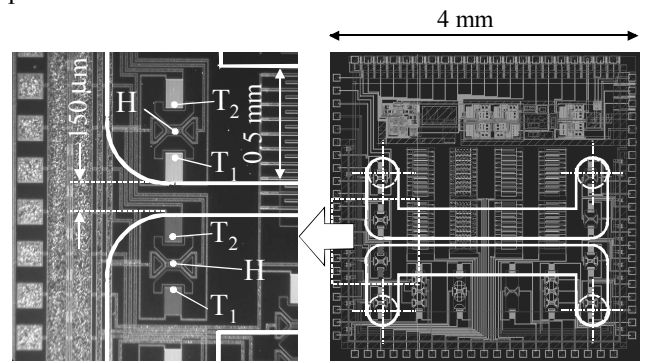

Fig. 4. Layout of the designed chip with the channel path.

The micrograph on the left shows an enlargement of the chip at the point the sensing structures are placed. The thermopiles ($T_1$ and $T_2$) and heater (H) are indicated for both structures. The small spacing between the sensing structures imposed to reduce the thickness of the wall separating the two trenches, requiring a fairly good precision in the mechanical fabrication of the adapter. Another unwanted consequence of the structure position and orientation was the necessity to place them practically inside a 90 degrees curve of the flow channel. Considering that the type of sensing structure used in this work is sensitive only to the gas velocity component along the thermopile-heater direction, a sensitivity reduction can be expected with respect to the case of Fig. 3 (b) where both structures are oriented along the gas streamlines.

### III   DEVICE FABRICATION

The chip was designed with the BCD6 (Bipolar-CMOS-DMOS) process of STMicroelectronics. Thermal insulation of the heaters and thermopile membranes from the substrate was obtained by suspending them on a cavity etched into the bulk silicon by means of a front side micromachining procedure. To this aim, proper openings are previously opened into the dielectric layers in order to access the bare silicon. The opening geometry was defined by means of a 1 µm resolution photolithographic step. This operation is facilitated by the presence of passivation openings purposely introduced in the layout design [12]. The substrate silicon is then etched through the openings by means of a TMAH based anisotropic etching performed at 90 ºC, two steps of 40 minutes each, renewing the solution after the first step. The described post-processing approach requires much shorter etching times than the alternative of removing the silicon by means of back-side micromachining.

The dies were glued to 28 pin DIL ceramic packages by means of epoxy resin and bonding of the pads to the package pins is performed. The chip includes also a few electronic blocks that were required to control the chip temperature. This possibility was exploited to optimize the adhesion of the chip surface to the plastic adapter. The adapter was fabricated using a precision milling machine (VHF CAM 100) controlled by a personal computer. The material for the adapter material was chosen considering characteristic of transparency, for alignment purposes, and low glass transition temperature, to facilitate the sealing procedure. Polymethylmethacrylate (PMMA) was chosen as a material for the adapter, considering also its greater hardness with respect to similar plastics [12]. Hardness was essential to allow fine milling of trenches with the small separation shown in Fig. 4. Alignment of the adapter to the chip was performed using a guide, schematically shown in Fig. 5(a), consisting in a thin PMMA bar with an opening of the same dimensions of the adapter front face. Firstly, the guide is leaned on the package surface by means of a micrometric displacement stage and coarsely aligned to the chip in order to allow the adapter to be inserted with no risk of hitting the bonding wires. Then, fine alignment, aimed to place the sensing structure into the channels, is performed with the adapter inside the guide. After the alignment, the guide was glued to the package surface by means of an epoxy resin.

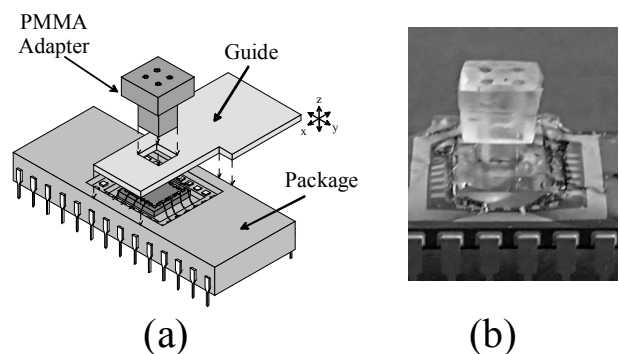

Fig. 5. (a) Perspective view illustrating the device assembling and (b) Photograph of the final device.

It is important to observe that this manual procedure can avoided in an industrial process, where it is possible to glue all the chips in a precise position of the package chamber and then simply align the guide to reference marks on the package itself.

At this stage it is not possible to obtain a leak proof connection of the adapter to the chip surface due to the uneven characteristic of the latter. In particular, protrusions of more than 3 µm, deriving from the thick upper metal layers, were measured with a stylus profilometer. Sealing of the flow channels was then obtained maintaining the chip at the PMMA glass transition temperature (110 ºC) for five minutes while applying a force of about 5 N between the chip itself and the adapter. A $\Delta V_{BE}$ substrate temperature sensor and a series of power DMOS present on the chip were used to maintain a constant temperature for the duration of the whole sealing procedure. Due to the poor adhesion of PMMA on silicon and silicon dioxide, the structure obtained in this way, though being leak proof, is rather fragile. A robust structure is obtained by carefully pouring epoxy resin inside the package chip chamber through holes in the guide. It is worth observing that, without the thermal sealing procedure, the resin slipped between the chip and the adapter, partially filling the trenches. The final structure of a two channel flow meter is shown in Fig.5(b).

### IV   EXPERIMENTAL RESULTS

Characterization of the devices was performed in nitrogen flow by means of a reference line, equipped with a mass flow controller (MKS 1179B) with a 10 sccm (standard cubic centimeters per minute) full scale range. The gas flow was applied to the devices by inserting a stainless steel pipe of 0.7 mm external diameter (syringe needle), into each one of the four upper orifices, in turn. Shifting the needle to one orifice to the other allowed to test the two channel in both flow directions. Before each measurement run, the needle was sealed to the orifice by means of silicone glue, which was easily removed after the measurement run in order to remove the needle and insert it in the next orifice. The thermopile output signal was read by means of a purposely built interface board, including a low noise instrumentation amplifier (AD 620), a second order 10 Hz low pass filter and a constant voltage source to bias the sensor heaters. In all the experiments the heaters were supplied by a 2 V constant voltage, corresponding to a power of 4 mW for each sensor.

The response of the two channel sensor is presented in Fig. 6. In details, Fig. 6(a) shows the thermopile output voltage of the channel with the PN structure, while Fig. 6(b) refers to the channel with the PA structure. A significant asymmetry with respect to flow reversal is visible for both channels. To better put in evidence this phenomenon, the linear fit of the data at flow rates less than 1 sccm has been drawn separately for the positive and negative branches of the curves. Inspection of Fig. 6 reveals that the asymmetry is much less important at small flow rates: it can then be ascribed to a non linearity effect, more evident for negative flows. Another aspect is the different sensitivity of the two channels, indicated by the slope of the linear fit reported in the figure distinctly for the negative and positive branch.

To further investigate the actual role of the channel geometry to determine the observed response features, experiments have also been performed with single channel devices. The result for the PA and PN structures is shown in Fig. 7(a) and 7(b), respectively. Note that, in this case, the two thermopile are included in the same channel. The fact that the PN structure still presents a sensitivity nearly 50 % higher than the PA one, clearly indicate that this is an intrinsic property of the thermopile type. It is more important to observe that the single channel sensor is generally more sensitive (almost one order of magnitude) and its response is also much more symmetric. This is a clear consequence of the fact that, for the two channel sensor, it has been necessary to place the sensing structures practically inside a channel elbow. The oblique gas velocity direction with respect to the thermopile-heater line may be the reason of the lower sensitivity of the double channel configuration.

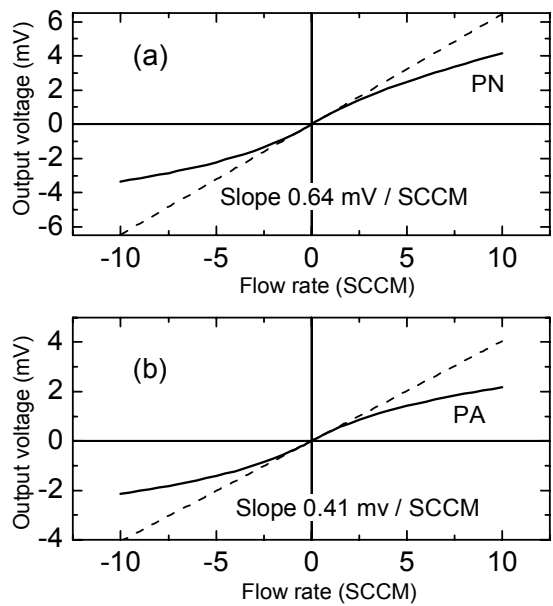

Fig. 7. Response of the two sensing structures placed inside the same L-shaped channel.

The asymmetry, instead, is probably a consequence of turbulences that may arise even at very low Reynolds numbers in the proximity of channel sharp bends.

Finally, it was important to test the cross-sensitivity between the two measurement channels. To this aim, flow rates up to 50 sccm have been injected into one channel while measuring the output signal of the other one. In all the experiments no measurable cross-talking was detected, confirming the excellent sealing of the channels.

## V    CONCLUSIONS

A previously presented technique for conveying gas flows to selected areas of a chip has been improved to allow multiple flow sensing with a single standard silicon chip. The fact that the chip was not designed to this purpose, prevented the placement of the structures into straight segments of the flow channels. As expected, this resulted in sensitivity reduction and increased response asymmetry. Reasonably, the design of a new chip with sensing structures with more favorable placement and orientation would eliminate this problem. However, the effective separation of the two flows, even in this critical configuration, demonstrated that the proposed technique can be successfully applied to probe adjacent areas of standard chips by distinct fluid flows. This suggests that the proposed inexpensive packaging approach can be applied also to the detection of liquid flows, where leakage towards critical chip areas (e.g. bonding pads), has to be strictly avoided.

ACKNOWLEDGMENT

The authors wish to thank the R&D group of STMicroelectronics in Cornaredo, Milan, for fabricating the chips used in the experiments described in this work.

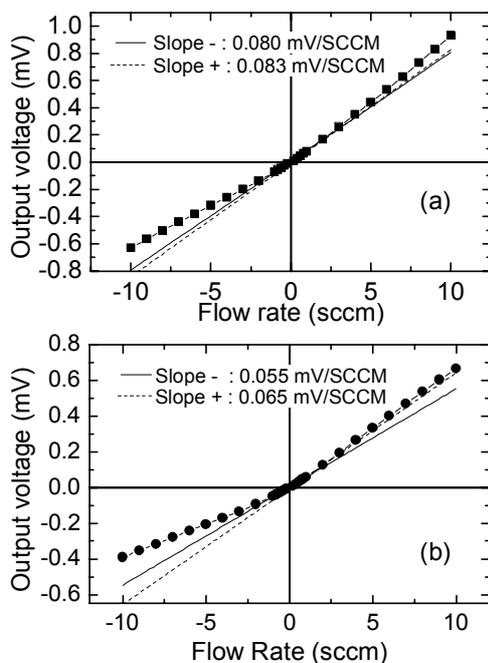

Fig. 6. Response of the two two-channel sensor after offset removal: (a) channel equipped with the PN structure, (b) channel with the PA structure. The linear fits are calculated for flow rates smaller than 0.8 SCCM, distinctly for the negative and positive branches.